\documentclass[epj]{svjour}
\usepackage{graphicx}

\begin{document}
\title{Entrance channel dependence in compound nuclear reactions with 
loosely bound nuclei}
\author{S.Adhikari\inst{1}, C. Samanta\inst{1,2}, C.Basu\inst{1}, 
S. Ray\inst{3}, A. Chatterjee\inst{4} \and S. Kailas\inst{4}
\thanks{Conference presenter Rituparna Kanungo}%
}                     
%
%
\institute{Saha Institute of Nuclear Physics, 1/AF Bidhan nagar, 
Kolkata - 700 064, India 
\and Physics Department, Virginia Commonwealth University, Richmond, 
VA 23284, USA
\and Dept. of Phys., University of Kalyani, Kalyani,  West Bengal - 741 235, India
\and Nuclear Physics Division, BARC, Mumbai - 400 085, India}
\date{Received: 15 April, 2004 / Revised version: date}
%
\abstract{
The measurement of light charged particles evaporated from the reaction
$^{6,7}$Li + $^6$Li has been carried out at extreme backward angle in the 
energy range 14 - 20 MeV. Calculations from the code ALICE91 show that the 
symmetry of the target-projectile combination and the choice of level
density parameter play important roles in explaining the evaporation spectra 
for these light particle systems. In above barrier energy region the fusion 
cross-section is not suppressed for these loosely bound nuclei.
\PACS{
      {25.70.Gh}{Compound nucleus}   \and
      {25.70.-z}{Low and intermediate energy heavy ion reactions}
     } 
} 
\authorrunning {S.Adhikari et al.}
\titlerunning {Entrance channel dependence ....}
\maketitle
\section{Introduction}
\label{intro}
Study of fusion reactions with loosely bound stable nuclei like $^{6,7}$Li, $^9$Be 
etc. have gained importance in recent times as they provide a good analogue 
to investigations with halo nuclei. The effect of low break-up threshold
of loosely bound nuclei on fusion reactions  
are not well understood \cite{ko98,tr02,be03,ta97}. Most of the recent experiments
with loosely bound nuclei investigate the behaviour of fusion excitation functions
both in the above and below barrier regions.
There are however fewer attempts to study the evaporation of light charged particles
involving the reaction of such nuclei \cite{le90,ka90}. 
In this work we report the inclusive measurement of $\alpha$ 
particles emitted in the reactions of $^{6,7}$Li projectile on $^6$Li target at extreme backward angle for a range of 
energies above the coulomb barrier. Statistical model calculations reproduce the
experimental $\alpha$ spectra (from other published works) nicely when emitted 
from a Compound nucleus (CN) formed from an asymmetric target projectile
combination. However in our case where the target-projectile combination is 
nearly symmetric, a large deformation (in terms of the 
rotating liquid drop model \cite{co74}) along with a structure dependent level 
density parameter is required to properly explain the observed spectra.

\section{Experiment}
\label{sec:1}
The experiment was performed using $^{6,7}$Li beams from the 14UD BARC-TIFR 
Pelletron Accelerator Facility at Mumbai, in the laboratory energy range 14 to 
20 MeV. A 4 mg/ cm$^2$ thick rolled $^6$Li target was used. Only light
 charged particles were detected. For particle identification standard two 
element $\Delta$E-E 
telescopes with silicon surface barrier ($\Delta$E = 10$\mu$m) and Si(Li) detectors 
(E=300 $\mu$m) were used. This telescope was placed at 175$^o$ to detect $\alpha$
particles.
 The beam current was kept between 1-20 nA. 
Standard electronics and CAMAC based data acquisition system were used. Energy 
calibration was done using $^7$Li elastic scattering data on Au and 
mylar targets each of thickness 500$\mu$g/cm$^2$.

\section{Results and discussions}
\label{sec:2}
Fig.1(a) and (b) show the inclusive $\alpha$-spectrum measured at 175$^o$
from the reaction $^7$Li+$^6$Li at energies E($^7$Li)= 14 and 16 MeV 
respectively. 
The experimental spectra in general consists of a continuum 
part followed by some discrete peaks at higher energy. 
The discrete peaks are identified as due to $\alpha$ emission from $^{13}$C 
and $^{23}$Na (formed due to oxidation of $^6$Li target) compound
 nuclei. As $^{6,7}$Li are loosely bound nuclei, the continuum part may 
contain contributions from both break-up and compound nuclear reactions. 
However, the contribution of $\alpha$-emission from break-up process at this extreme 
backward angle is expected to be negligible. To evaluate the continuum part
we use the statistical model code ALICE91 \cite{bl91}.  
The dashed lines in Fig.1 indicate calculations with the Fermi 
gas level density parameter ($a=A/9$) assuming a spherical compound nucleus.  
As can be seen, the calculations grossly overpredict the experimental data. 
It is well known that the level density parameter $a$ strongly influences 
the level density and hence the higher energy part of the evaporation spectra. 
We have found that in our case arbitrary change of $a$ parameter does not help 
to improve the spectra, except for some change in slope. Instead of resorting 
to arbitrary adjustment of the parameters in the statistical model we try 
to consider a deformation in the excited CN as in the works \cite{go00,bh01} 
for reactions with heavier nuclei. 
The dotted lines show the results of this calculation with the same level 
density parameter. The rotational energy is evaluated in terms of the 
Rotating Liquid Drop Model deformations \cite{co74}. These new calculations  are now much 
reduced in comparison to the calculations assuming a spherical CN but they 
still overpredict the data. In the Fermi gas model {\it a/A} 
is simply a constant. However there are shell effects in {\it a} especially 
near the magic nucleon numbers. Therefore, instead of trying to adjust the 
parameter {\it a}, we 
now adopt the Gilbert Cameron prescription \cite{gi65} for the shell dependent level 
density parameter. The solid lines are calculations using shell corrected 
Gilbert-Cameron level density parameter and a deformed CN. This calculation 
agrees with the experimental data more satisfactorily. Similar results were 
observed for the reactions with $^6$Li beam which will be discussed 
elsewhere.   
%
\begin{figure*}
\includegraphics{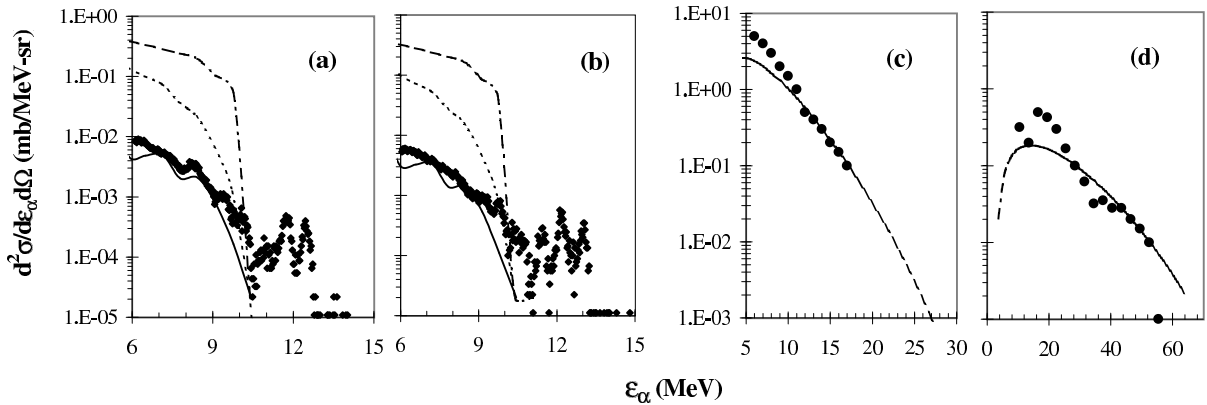}
\caption{{\bf (a), (b)} Inclusive $\alpha$-spectrum measured at 175$^o$  
from the reaction $^7$Li + $^6$Li at energies E($^7$Li)=14 and 16 MeV 
respectively. Inclusive $\alpha$-spectrum from {\bf (c)} 
$^{28}$Si + $^6$Li scattering at E($^6$Li)=36 MeV, $\theta$=155$^o$ and 
{\bf (d)} n + $^{12}$C scattering at E(n)=72.8 MeV, $\theta$=40$^o$.
Calculations (solid, dashed and dotted lines) are explained in the text.}  
\label{fig:1}       
\end{figure*}
%
 In order to verify the effect of target-projectile symmetry on the statistical
calculations we have 
reanalyzed the published experimental data for $^{12}$C(n,$\alpha$) \cite{sl00}
and  $^{28}$Si($^6$Li,$\alpha$) \cite{ka90} shown in Fig.1(c) and (d). In the 
reaction 
$^{12}$C(n,$\alpha$) the compound nucleus is $^{13}$C, which is same as in our 
experiment but populated by an asymmetric combination of target and projectile.
However, the excitation energy is much higher (72.15 MeV) (lower energy 
data is not available for this system).In case of 
$^{28}$Si($^6$Li,$\alpha$) the $\alpha$-particles were detected at backward 
angle (155$^o$) and the excitation of the compound nucleus was 46.68 MeV. 
The excitation of the compound nucleus in our case is close 
to this value (32.34 to 35.1 MeV).  ALICE91 calculations (dotted line) using 
Fermi gas level density parameter ($a=A/9$) in the Weisskopf-Ewing approximation
(without any deformation) 
and comparison to the observed data is shown in Fig.1(c) and (d). The calculation 
reproduces the experimental data satisfactorily considering a spherical compound 
nucleus. 

In summary, the measurement of light charged particles evaporated from  
$^{6,7}$Li + $^6$Li has been carried out at extreme backward angle in the 
energy range 14 - 20 MeV. Calculations considering 
a deformed compound nucleus and  shell-corrected Gilbert-Cameron level 
density parameter
agree well with the experimental data. Interestingly, statistical model  
calculations require the excited compound nucleus to be deformed for 
$^{6,7}$Li+$^6$Li 
reaction, but spherical for asymmetric n+$^{12}$C target-projectile 
combination. This 
indicates some target-projectile dependence for the light particle evaporation 
spectra. This phenomena though known for heavier systems has not 
been reported earlier for such light loosely bound nuclei. For further 
verification of this phenomenon, additional experimental data leading to 
the same compound nucleus, excitation energy and angular momentum are needed 
in both the symmetric and asymmetric channels.

\noindent Authors thank V. Tripathi, K. Mahata, K. Ramachandran and 
Pelletron personnel for their generous help during experiment and grant no. 2000/37/30/BRNS 
for funding. 

%
%
%

\end{document}